\documentclass[sigconf]{acmart}

\usepackage{multirow}
\usepackage{booktabs} % For formal tables
\usepackage{subfig}
\usepackage{xspace}

\graphicspath{{Figures/}}

\newcommand{\ie}{\emph{i.e.,}\xspace}
\newcommand{\eg}{\emph{e.g.,}\xspace}
\newcommand{\etal}{\emph{et al.}\xspace}
\newcommand{\paratitle}[1]{\vspace{1ex}\noindent \textbf{#1}}

\setcopyright{rightsretained}
\acmDOI{}

\acmISBN{}

\acmConference[]{Conference Submission}{Month 2018/9}{}
\acmYear{2018}
\copyrightyear{2018}

\acmPrice{15.00}

\begin{document}
\title{Next Item Recommendation with Self-Attention}
% \title{Self-Attentive Metric Learning for Sequential Recommendation}

\author{Shuai Zhang}
\orcid{1234-5678-9012}
\affiliation{%
  \institution{UNSW and Data61, CSIRO}
  \city{Sydney}
  \state{NSW}
  \postcode{2052}
 \country{Australia}
}
\email{shuai.zhang@student.unsw.edu.au}

\author{Yi Tay}
\affiliation{%
  \institution{Nanyang Technological University}
  \country{Singapore}
}
\email{ytay017@e.ntu.edu.sg}

\author{Lina Yao}
\affiliation{%
  \institution{University of New South Wales}
  %\streetaddress{K17, CSE, UNSW}
  \city{Sydney}
  \state{NSW}
  \postcode{2052}
  \country{Australia}
}
\email{lina.yao@unsw.edu.au}

% \author{Xing Xie}
% \affiliation{
%   \institution{Brookhaven Laboratories}
%   \streetaddress{P.O. Box 5000}}
% \email{lleipuner@researchlabs.org}

\author{Aixin Sun}
\affiliation{%
  \institution{Nanyang Technological University}
  \country{Singapore}
}
\email{axsun@ntu.edu.sg}

% \renewcommand{\shortauthors}{Zhang et al.}

\begin{abstract}
% Predicting a user's next interaction, given their past behaviour forms the crux of sequential recommender systems. This paper proposes a new neural architecture for the sequential recommendation task. We demonstrate the utility of explicit item-item interactions in the sequential recommender task. Our intuition manifests in the form of an attention-based model, relying on item-item attention matrices to capture these intra-user item interactions. Our method explicitly invokes interactions between all items in a user's sequence window with self-attention. Based on these interactions, we learn the relative importance of each item in user's interaction sequence to accurately infer user's intents. Overall, our model takes the form of a metric learning framework and can be trained end-to-end with standard ranking objectives. We conduct extensive experiments on twelve publicly available datasets. Our proposed approach outperforms the current state-of-the-art models on all datasets with an average improvement of $13.96\%$, with margins ranging from $2.82\%$ to $45.03\%$ depending on the dataset.

In this paper, we propose a novel sequence-aware recommendation model. Our model utilizes self-attention mechanism to infer the item-item relationship from user's historical interactions. With self-attention, it is able to estimate the relative weights of each item in user interaction trajectories to learn better representations for user's transient interests.  The model is finally trained in a metric learning framework, taking both short-term and long-term intentions into consideration. Experiments on a wide range of datasets on different domains demonstrate that our approach outperforms the state-of-the-art by a wide margin.

\end{abstract}


\maketitle

%============================================
\section{Introduction}
%============================================

Anticipating a user's next interaction lives at the heart of making personalized recommendations. The importance of such systems cannot be overstated, especially given the ever growing amount of data and choices that consumers are faced with each day~\cite{quadrana2018sequence}. Across a diverse plethora of domains, a wealth of historical interaction data exists, \eg click logs, purchase histories, views etc., which have, across the years, enabled many highly effective recommender systems.

Exploiting historical data to make future predictions have been the cornerstone of many machine learning based recommender systems. After all, it is both imperative and intuitive that a user's past interactions are generally predictive of their next. To this end, many works have leveraged upon this structural co-occurrence, along with the rich sequential patterns, to make informed decisions. Our work is concerned with building highly effective sequential recommender systems by leveraging these auto-regressive tendencies.

In the recent years, neural models such as recurrent neural network (RNN)/convolutional neural network (CNN) are popular choices for the problem at hand~\cite{goodfellow2016deep,lecun2015deep}. In recurrent models, the interactions between consecutive items are captured by a recurrent matrix and long-term dependencies are persisted in the recurrent memory while reading. On the other hand, convolution implicitly captures interactions by sliding parameterized transformations across the input sequence~\cite{gehring2017convolutional}. However, when applied to recommendation, both models suffer from a shortcoming. That is, they fail to \textbf{explicitly} capture item-item interactions\footnote{In RNNs, this is captured via memory persistence. While in CNNs, this is only weakly captured by the sliding-window concatenated transformations. In both cases, there is no explicit interaction.} across the entire user history. The motivation for modeling item-item relationships within a user's context history is intuitive, as it is more often than not, crucial to understand fine-grained relationships between individual item pairs instead of simply glossing over them. All in all, we hypothesize that providing an inductive bias for our models would lead to improve representation quality, eventually resulting in a significant performance improvement within the context of sequential recommender systems.

To this end, this paper proposes a new neural sequential recommender system where sequential representations are learned via modeling not only consecutive items but across \textbf{all user interactions} in the current window. As such our model can be considered as a `local-global' approach. Overall, our intuition manifests in the form of an attention-based neural model that explicitly invokes item-item interactions across the entire user's historical transaction sequence. This not only enables us to learn global/long-range representations, but also short-term information between $k$-consecutive items. Based on this self-matching matrix, we learn to attend over the interaction sequence to select the most relevant items to form the final user representation. Our  experiments show that the proposed model outperforms the state-of-the-art sequential recommendation models by a wide margin, demonstrating the effectiveness of not only modeling local dependencies but also going global.

Our model takes the form of a metric learning framework in which the distance between the self-attended representation of a user and the prospective (golden) item is drawn closer during training. To the best of our knowledge, this is the first proposed attention-based metric learning approach in the context of sequential recommendation. To recapitulate, the prime contributions of this work are as follows:
\begin{itemize}
    \item We propose a novel framework for sequential recommendation task. Our model combines self-attention network with metric embedding to model user temporary as well as long-lasting intents.
    \item Our proposed framework demonstrates the utility of explicit item-item relationships during sequence modeling by achieving state-of-the-art performance across \textbf{twelve} well-established benchmark datasets. Our proposed model outperforms the current state-of-the-art models (\eg Caser and TransRec) on all datasets by margins ranging from $2.82\%$ to $45.03\%$ (and $13.96\%$ on average) in terms of standard retrieval metrics.
    \item We conduct extensive hyper-parameter and ablation studies. We study the impacts of various key hyper-parameters and model architectures  on model performance. We also provide a qualitative visualisation of the learned attention matrices.
    \end{itemize}

%============================================
\section{Related Work}
%============================================
In this section, we briefly review the related works from three perspectives: sequence-aware recommender systems, deep neural network models for recommendations, and neural attention models.

%============================================
\subsection{Sequence-aware Recommender Systems}
%============================================

In many real-world applications, user-item interactions are recorded over time with associated timestamps. The accumulated data enables modelling temporal dynamics and provides evidence for user preference refinement~\cite{koren2009collaborative,quadrana2018sequence,Rendle:2010:FPM:1772690.1772773,he2016fusing,chen2018sequential}. Koren \etal~\cite{koren2009collaborative} propose treating user and item biases as a function that changes over time, to model both item transient popularity and user temporal inclinations. Xiong \etal~\cite{xiong2010temporal} introduce additional factors for time and build a Bayesian probabilistic tensor factorization approach to model time drifting. Wu \etal~\cite{Wu:2017:RRN:3018661.3018689} use recurrent neural network to model the temporal evolution of ratings. Nonetheless, these methods are specifically designed for the rating prediction task.

To generate personalized ranking lists, Rendle \etal~\cite{Rendle:2010:FPM:1772690.1772773} propose combining matrix factorization with Markov chains for next-basket recommendation.  Matrix factorization can capture user's general preference while Markov chain is used to model the sequential behavior. He \etal~\cite{he2016fusing} describe a sequential recommendation approach which fuses similarity based methods with Markov chain. Apart from Markov Chain, metric embedding has also shown to perform well on sequence-aware recommendation. Feng \etal~\cite{feng2015personalized} introduce a Point-of-Interest recommender with metric embedding to model personalized check-in sequences. Then, He \etal improve this model by introducing the idea of translating embedding~\cite{bordes2013translating,he2017translation}. This approach views user as the relational vector acting as the junction between items.  The major advantage of using metric embedding instead of matrix factorization is that it satisfies the transitive property of inequality states~\cite{hsieh2017collaborative,tay2018latent,zhangmetric}.

%============================================
\subsection{Deep Neural Network for Recommendation}
%============================================

Deep learning has been revolutionizing the recommender systems. In both industry and academia, the achievements of deep learning based recommender systems are inspiring and enlightening~\cite{zhang2017deep}. Former studies show that a various of deep learning techniques can be applied on many recommendation tasks.  For instance, multilayer perception can be used to introduce nonlinearity in modelling user item relationship~\cite{he2017neural,wang2017item,Zhang:2017:JRL:3132847.3132892,zhang2018neurec}. Convolutional neural network can be used to extract features from textual, visual and audio information sources of items and users~\cite{Wang:2017:YIR:3038912.3052638,van2013deep,he2018outer,he2016ups,he2016vbpr,Chen:2017:PKF:3077136.3080776}. Autoencoder learns salient feature representations from side information to enhance recommendation quality~\cite{wang2015collaborative,Zhang:2017:AEH:3077136.3080689,zhang2017hybrid}. Recurrent neural network  is capable of modelling the temporal dynamics~\cite{Wu:2017:RRN:3018661.3018689,hidasi2015session,Jannach:2017:RNN:3109859.3109872}.

Among the different recommendation tasks, session-based recommendation looks similar to our task but with a fundamental difference. In session based recommender,  user identification is usually unknown, thus the model structure and learning procedure are divergent. The flexibility of deep neural network makes it possible to combine different neural networks together to form more powerful hybrid recommenders. The successful applications mainly attribute to the nonlinearity, powerful representation learning, and sequence modelling capability of deep neural networks.

Specific to sequential recommendation, many deep neural network models have been proposed. Wang \etal~\cite{Wang:2015:LHR:2766462.2767694} introduce a two-layer network named hierarchical representation model, to capture both user general preference and sequential behavior. However, this model does not incorporate any non-linear transformation.  More recently, Tang \etal~\cite{Tang:2018:PTS:3159652.3159656} propose a convolutional sequence modelling method to learn user's transient trajectory with horizontal and vertical convolutional layer. This model achieves better performance than RNN based approaches~\cite{DBLP:journals/corr/HidasiKBT15,yu2016dynamic}. On the other hand, this model falls short of dealing with recommendation on sparse dataset. In general, CNN and RNN need to learn from a large amount of data to come up with meaningful results, and data sparsity makes the model learning rather difficult.

The main difference between our work and existing approaches is the use of self-attention mechanism.  Proposed in  Transformer~\cite{vaswani2017attention}, self-attention mechanism  brings the benefits of automatically learning the importance of past behaviors. Furthermore, we design a structure to combine metric learning with self-attention to consider user inclinations, both short-term and long-term.

%============================================
\subsection{Neural Attention Models}
%============================================

The neural attention mechanism shares similar intuition with that of the visual attention found in humans. It learns to pay attention to only the most important parts of the target, and has been widely employed across a number of applications \eg natural language processing and computer vision. Standard vanilla attention mechanism can be integrated into CNN and RNN to overcome their shortcomings. Specifically, attention mechanism makes it easy to memorize very long-range dependencies in RNN, and helps CNN to concentrate on important parts of inputs. Several recent studies also investigated its capability in recommendation tasks such as hashtag recommendation~\cite{gong2016hashtag}, one-class recommendation~\cite{chen2017attentive,8352808,tay2018multi,tay2018couplenet}, and session based recommendation~\cite{li2017neural}.

Our work is concerned with a new concept known as `self-attention', or  `intra-attention'. Different from the standard vanilla attention, self-attention focuses on co-learning and self-matching of two sequences whereby the attention weights of one sequence is conditioned on the other sequence, and vice versa. It has only started to gain exposure due to its recent successful application on machine translation~\cite{vaswani2017attention}. It can replace RNN and CNN in  sequence learning, achieving better accuracy with lower computation complexity. In this work, we utilize self-attention to model dependencies and  importance of user short term behavior patterns. Note that, the usage of
self-attention in the context of recommender systems is far from straightforward, substantially contributing to the overall novelty of our work.

%============================================
\section{The Proposed Model: AttRec}
%============================================

We now present the proposed self-attentive sequential recommendation model, named \textbf{AttRec}. Our model consists of a \textit{self-attention module} to model user short-term intent, and a \textit{collaborative metric learning} component to model user long-term preference. Next, we formally define the task of sequential recommendation.

%============================================
\subsection{Sequential Recommendation}
%============================================

Sequential recommendation is very different from traditional one-class collaborative filtering recommendation.  Let $\mathcal{U}$ be a set of users and $\mathcal{I}$ be a set of items, where $|\mathcal{U}|=M$ and $|\mathcal{I}|=N$. We use $\mathcal{H}^u = (\mathcal{H}^u_1,\cdots, \mathcal{H}^u_{|\mathcal{H}^u|})$ to denote a sequence of items in chronological order that user $u$ has interacted with before, where $\mathcal{H}^u \sqsubseteq \mathcal{I}$. The objective of sequential recommendation is to predict the next items that the user will interact with, given her former consumption trajectory.

\begin{figure}
    \centering
    \includegraphics[width=0.48\textwidth]{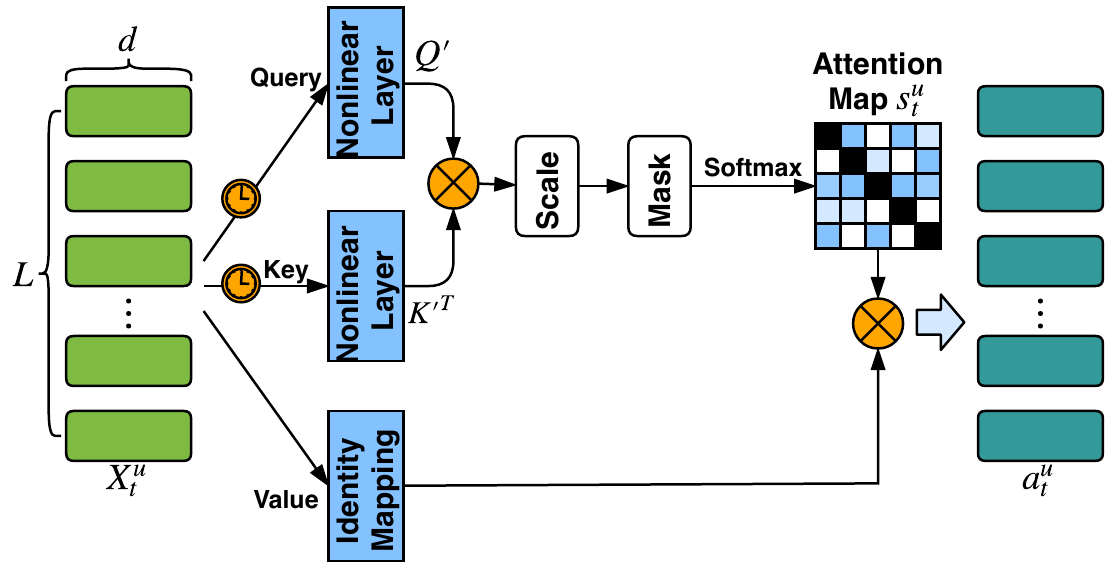}
    \caption{Illustration of the self-attention module. The input  is the embedding matrix of the latest interacted $L$ items, and the output is the self-attentive representations.}
    \label{fig:selfatt}
\end{figure}

%============================================
\subsection{Short-Term Intents Modelling with Self-Attention}
%============================================

User recent interactions reflect user's demands or intents in a near term. Modelling user short-term interaction therefore is an important task for better understanding of user's temporal preferences. To this end, we propose leveraging the recent success of self-attention mechanism in capturing sequential patterns, and use it to model user's recent interaction trail. Figure~\ref{fig:selfatt} illustrates the proposed self-attention module in our method.

%============================================
\paratitle{Self-Attention Module.} Self-attention is an special case of the attention mechanism and has been successfully applied to a variety of tasks. It refines the representation by matching a single sequence against itself. Unlike basic attention that learns representations with limited knowledge of the whole context, self-attention can keep the contextual sequential information and capture the relationships between elements in the sequence, regardless of their distance. Here, we apply self-attention to attend user's past behaviours.

The building block of self-attention is scaled dot-product attention. The input of the attention module consists of \emph{query}, \emph{key}, and \emph{value}. The output of attention is a weighted sum of the \emph{value}, where the weight matrix, or affinity matrix, is determined by \emph{query} and its corresponding \emph{key}. In our context, all of these three components (\ie \emph{query}, \emph{key}, and \emph{value}) are the same and composed from user recent interaction histories (see Figure~\ref{fig:selfatt}).

Suppose user's short-term intents can be derived from her recent $L$ (\eg 5, 10) interactions. Assuming each item can be represented with a $d$-dimension embedding vector. Let $X \in \mathcal{R}^{N \times d}$ denote the embedding representations for the whole item set. The latest $L$ items (\ie from item $t-L + 1$ to item $t$) are stacked together in sequence to get the following matrix.
\begin{equation}
    X^u_{t} = \begin{bmatrix}
    X_{(t-L+ 1)1} & X_{(t-L + 1)2} &\dots  & X_{(t-L+ 1)d} \\
    \vdots & \vdots & \vdots & \vdots \\
    X_{(t-1)1} & X_{(t-1)2} &  \dots  & X_{(t-1)d} \\
    X_{t1} & X_{t2}  & \dots  & X_{td}
\end{bmatrix}
\end{equation}
Here, the latest $L$ items is a subset of $\mathcal{H}^u$. \emph{Query}, \emph{key},  and \emph{value} for user $u$ at time step $t$ in the self-attention model equal to $ X^u_{t}$.

First, we project \emph{query} and \emph{key} to the same space through nonlinear transformation with shared parameters.
\begin{align}
    Q' = ReLU(X^u_{t} W_Q) \\
    K' = ReLU(X^u_{t} W_K)
\end{align}
where $W_Q \in \mathcal{R}^{d \times d}  =  W_K \in \mathcal{R}^{d \times d}$ are weight matrices for \emph{query} and \emph{key} respectively. \textit{ReLU} is used as the activation function, to introduce some non-linearity to the learned attention. Then, the affinity matrix is calculated as follows:
\begin{equation}
    s^u_t = \text{softmax}(\frac{Q'K'^T}{\sqrt{d}})
\end{equation}
The output is a $L \times L$ affinity matrix (or attention map) which indicates the similarity among $L$ items. Note that the $\sqrt{d}$ is used to scale the dot product attention. As in our case, $d$ is usually set to a large value (\eg 100), so the scaling factor could reduce the extremely small gradients effect. A masking operation (which masks the diagonal of the affinity matrix) is applied before the softmax, to avoid high matching scores between identical vectors of \emph{query} and \emph{key}.

Second, we keep the \emph{value} equals to $X^u_{t}$ unchanged. Unlike in other cases~\cite{vaswani2017attention} where \emph{value} is usually mapped with linear transformations, we found that it is beneficial to use identity mapping in our model. In other application domains, the \emph{value} is usually pretrained feature embeddings such as word embedding or image features.  In our model, the \emph{value} is made up of parameters that need to be learned. Adding linear (or nonlinear) transformation will increase the difficulty in seeing the actual parameters. Note that \emph{query} and \emph{key} are used as auxiliary factors so that they are not as sensitive as  \emph{value} to transformations.

Finally, the affinity matrix and the \emph{value} are multiplied to form the final weighted output of the self-attention module.
\begin{equation}
    a^u_t = s^u_t X^u_{t}
\end{equation}
Here, the attentive output $a^u_t \in \mathcal{R}^{L \times d}$ can be viewed as user's short-term intent representations. In order to learn a single attentive representation, we take the mean embedding of the $L$ self-attention representations as  user temporal intent. Note that other aggregation operation (\eg  sum, max, and min)  can also be used and we will evaluate their effectiveness in our experiments.
\begin{equation}
    m^u_t = \frac{1}{L} \sum^{L}_{l=1} a^u_{tl}
\end{equation}

%============================================
\paratitle{Input Embedding with Time Signals.} The above attention model does not include  time signals. Without time sequential signals, the input  degrades to bag of embeddings and fails to retain the sequential patterns. Following the Transformer, we propose to furnish the \emph{query} and \emph{key} with time information by positional embeddings.  Here, we use a geometric sequence of timescales to add sinusoids of different frequencies to the input. The time embedding (TE) consists of two sinusoidal signals defined as follows.
\begin{align}
    TE(t, 2i) = sin(t/10000^{2i/d}) \\
    TE(t, 2i+1) = cos(t/10000^{2i/d})
\end{align}
Here, $t$ is the time step and $i$ is the dimension. The TE is simply added to \emph{query} and \emph{key} before the nonlinear transformation.

%============================================
\subsection{User Long-Term Preference Modelling}
%============================================

After modelling the short-term effects, it is beneficial to incorporate general tastes or long-term preference of users. Same as latent factor approach, we assign each user and each item a latent factor. Let $U \in \mathcal{R}^{M \times d}$ and $V \in \mathcal{R}^{N \times d}$ denote the latent factors of users and items. We could use dot product to model the user item interaction as in latent factor model. However, recent studies~\cite{hsieh2017collaborative,tay2018latent} suggest that dot product violate the important inequality property of metric function and will lead to sub-optimal solutions. To avoid this problem, we adopt the Euclidean distance to measure the closeness between item $i$ and user $u$.
\begin{equation}
    \parallel U_u - V_i \parallel^2_2
\end{equation}
The distance is expected to be small if user $u$ liked the item $i$, and  large otherwise.

\begin{figure}
    \centering
    \includegraphics[width=0.48\textwidth]{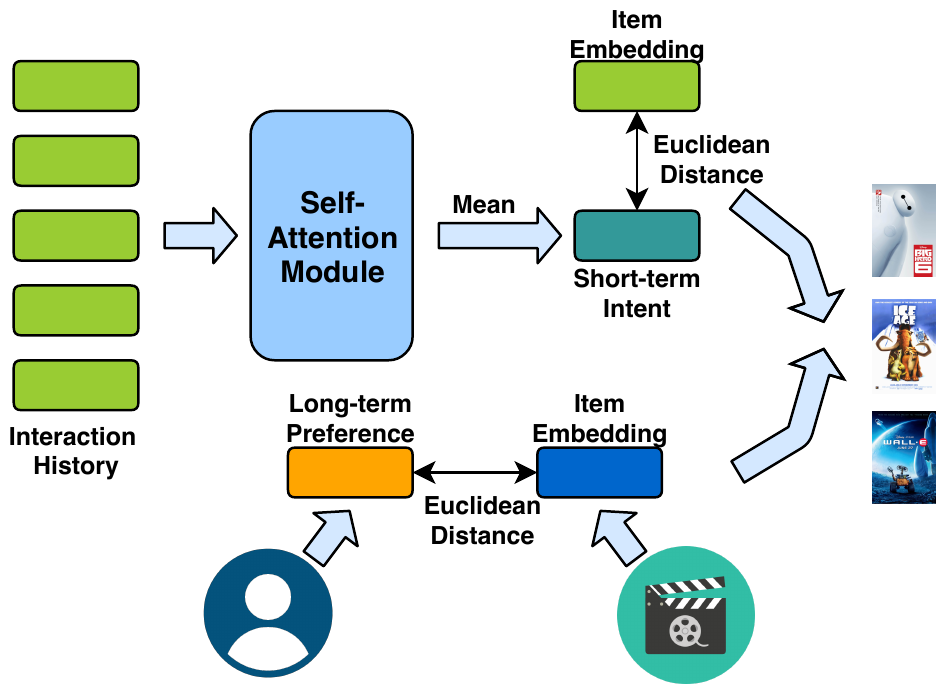}
    \caption{The architecture of the self-attentive metric learning approach for sequential recommendation. This model combine self-attention network with metric learning and considers both user's short-term and long-term preference.}
    \label{fig:arch}
\end{figure}

\begin{table*}
\caption{Statistics of the  datasets used in experiments.}
\begin{tabular}{c|rrrrrr}

\toprule
Dataset   & \#Users & \#Items & \#Interactions & Density & avg. \#actions per user & Time Interval \\
\midrule
ML-100K   & 943     & 1,682    &   100,000     &  6.30\%  &  106.04   &    Sept/1997 - Apr/1998 \\
ML-HetRec &  2,113 & 10,109&  855,598  &   4.01\%    & 404.92 &  Sept/1997 - Jan/2009   \\
ML-1M     &   6,040      &   3,706      &  1,000,209              &  4.46\%       &      1655.66   &     Apr/2000 - Mar/2003  \\
Android App  & 21,309 &  19,256&  358,877  &     0.087\%    &   16.84   &  Mar/2010 - Jul/2014 \\
Health / Care &    11,588     &     31,709    &   211,284  & 0.058\%   &     18.23 &   Dec/2000 - Jul/2014    \\
 Video Game  &    7,220     &  16,334       &  140,307  &     0.119\%    &   19.43  & Nov/1999 - Jul/2014 \\
   Tools / Home &    5,855     &     23,620    &   96,467            &      0.070\%   &     16.48   & Nov/1999 - Jul/2014        \\
   Digital Music &    2,893     &     13,183    &   64,320            &      0.169\%   &     22.23   &    May/1998 - Jul/2014        \\
   Garden &    1,036     &     4,900    &   15,576            &      0.307\%   &     15.03                    &  Apr/2000 - Jul/2014   \\
 Instant Video &    1,000     &     3,296    &   15,849   &  0.481\%   &     15.85                    &   Aug/2000 - Jul/2014         \\

  LastFM        &    951     &  12,113       &  83,382              &     0.724\%    &   87.68&  Aug/2005 - May/2011\\
  MovieTweetings     &   9,608      &   14,220      &  461,970              &  0.338\%       &      48.08   &  Mar/2013 - Dec/2016          \\
\bottomrule
\end{tabular}
\end{table*}

%============================================
\subsection{Model Learning}
%============================================

\paratitle{Objective Function.} Given the short-term attentive intents at time step $t$ and long-term preference, our task is to predict the item (denoted by $\mathcal{H}^u_{t+1}$) which the user will interact with at time step $t+1$.  To keep consistency, we adopt Euclidean distance to model both short-term and long-term effects, and use their sum as the final recommendation score.
\begin{equation}
    y^u_{t+1} =  \omega \parallel U_u - V_{\mathcal{H}^u_{t+1}} \parallel^2_2 + (1 -\omega)  \parallel m^u_t - X^u_{t+1} \parallel^2_2
    \label{eq:recscore}
\end{equation}
In the above equation, the first term denotes the long-term recommendation score between user $u$ and the next item $\mathcal{H}^u_{t+1}$, while the second term indicates the short-term recommendation score between user $u$ and its next item. Note that both $V_{\mathcal{H}^u_{t+1}}$ and $ X^u_{t+1}$ are the embedding vectors for the next item, but $V$ and $X$ are two different parameters. The final score is a weighted sum of them with the controlling factor $\omega$.

In some cases, we may want to predict the next several items instead of just one item. It enables our model to capture the skip behaviours in the sequence~\cite{Tang:2018:PTS:3159652.3159656}. Let $\mathcal{T}^+$ denote the next $T$ items that user liked in groundtruth. In this paper, we adopt a pairwise ranking method to learn the model parameters. Thus we need to sample $T$ negative items that the user does not interact with (or dislike) and denote this set by $\mathcal{T}^-$. Apparently  $\mathcal{T}^-$ is sampled from the set $\mathcal{I} \backslash \mathcal{T}^+$. To encourage discrimination between positive user item pair and negative user item pair, we use the following margin-based hinge loss.
\begin{equation}
\mathcal{L}(\Theta) = \sum_{(u, i) \in \mathcal{T}^+} \sum_{(u, j) \in \mathcal{T}^-} [y^u_i + \gamma - y^u_j]_+ + \lambda \parallel \Theta \parallel^2_2
\end{equation}
In the above equation, $\Theta = \{X, V, U, W_Q, W_K\}$ represents model parameters. $\gamma$ is the margin that separates the positive and negative pairs. We use $\ell_2$ loss to control the model complexity. Dropout can also be applied for nonlinear layer of the self-attention module. Because we use Euclidean distance in our method, for sparse datasets, we could also alternatively apply the norm clipping strategy to constrain $X, V, U$ in a unit Euclidean ball.
\begin{equation}
    \parallel X_* \parallel_2 \leq 1, \parallel V_* \parallel_2 \leq 1,
    \parallel U_* \parallel_2 \leq 1
\end{equation}
This regularization approach is useful for sparse dataset to alleviate the curse of dimensionality problem and prevent the data points from spreading too broadly.

%============================================
\paratitle{Optimization and Recommendation.}  We optimize the proposed approach with adaptive gradient algorithm~\cite{duchi2011adaptive} which could adapt the step size automatically; hence it reduces the efforts in learning rate tuning. In the recommendation stage, candidature items are ranked in ascending order based on the recommendation score computed by Equation~(\ref{eq:recscore}) and the top ranked items are recommended to users.

Since we optimize the model in a pairwise manner, the training time complexity for each epoch is $\mathcal{O}(\mathcal{K}d)$ where $K$ is the number of observed interactions. Note that $K \ll MN$ so that it can be trained efficiently. In the recommendation stage, we  compute the recommendation scores for all user item pairs at once and generate the ranking lists with efficient sort algorithms. As such, the whole process can run very fast.

Figure~\ref{fig:arch} illustrates the architecture of our model. It includes not only user transient intents but also long-lasting preference. Both are added up to generate the final recommendation lists. The former is inferred by self-attention network from recent actions and the whole system is constructed under the metric learning framework.

%============================================
\section{Experiments}
%============================================

We evaluate the proposed model on a wide spectrum of real world datasets. We then conduct  detailed ablation studies. In short, our experiments are designed to answer the following research questions:

\paratitle{RQ1:} Does the proposed self-attentive sequential recommendation model achieve state-of-the-art performance? Can it deal with sparse datasets?

\paratitle{RQ2:}  What is the effect of the key hyper-parameters? For example, the aggregation method and the length of sequence for mining short-term intents.

%============================================
\subsection{Datasets Descriptions}
%============================================

We conduct experiments on the following datasets. All of them include time-stamps of interactions.

\paratitle{MovieLens}. This is a popular benchmark dataset for evaluating the performance of recommendation models. We adopt three well-established versions in our experiments: Movielens 100K, Movielens HetRec and Movielens 1M\footnote{https://grouplens.org/datasets/movielens/}.

\paratitle{Amazon}. This is a user purchase and rating dataset collected from Amazon, a well-known e-commerce platform, by McAuley \etal~\cite{he2016ups,mcauley2015image}. In this work, we adopt 7 sub-categories: Android Apps, Health/Care, Video Game, Tools/Home, Digital Music, Garden and Instant Video, due to limited space.

\paratitle{LastFM}. This dataset contains user tag assignments collected from last.fm online music system\footnote{ http://www.lastfm.com}.

\paratitle{MovieTweetings}. It is obtained by scraping Twitter for well-structured tweets for movie ratings. This dataset is comparatively new and being updated\footnote{https://github.com/sidooms/MovieTweetings}. The subset we used was downloaded in December 2016.

For datasets with explicit ratings, we convert it to implicit feedback following early studies~\cite{he2017neural,tay2018latent}. For Amazon, lastFM and MovieTweetings, we perform a modest filtering similar to~\cite{Rendle:2010:FPM:1772690.1772773,Wang:2015:LHR:2766462.2767694,feng2015personalized,he2017translation,Tang:2018:PTS:3159652.3159656} to discard users with fewer than $10$ associated actions and remove cold-start items.  This is a common pre-process to reduce the noise of cold start issue as it is usually addressed separately~\cite{Tang:2018:PTS:3159652.3159656}. Detail statistics of the datasets are presented in Table 1.  It shows Movielens datasets are more dense than other datasets.

%============================================
\subsection{Evaluation Metrics}
%============================================

For each user, we use the most recent item for test and the second most recent item for hyper-parameter tuning. We evaluate the performance of all the models with hit ratio and mean reciprocal rank (MRR). Hit ratio measures the accuracy of the recommendation. We report the hit ratio with cut off value $50$, defined as follows:
\begin{equation}
    HR@50 = \frac{1}{M}\sum_{u\in \mathcal{U} } \textbf{1}(R_{u, g_u} \leq 50)
\end{equation}
Here, $g_u$ is the item that  user $u$ interacted with at the most recent time step. $R_{u, g_u}$ is the rank generated by the model for this groundtruth item. If a model ranks $g_u$ among the top 50, the indicator function will return 1, otherwise 0.

Mean Reciprocal Rank indicates how well the model rank the item. Intuitively, ranking the groundtruth item higher is more preferable in practice. The definition of MRR is as follows:
\begin{equation}
    MRR = \frac{1}{M}\sum_{u\in \mathcal{U} } \frac{1}{R_{u, g_u}}
\end{equation}
$R_{u, g_u}$ is the rank for groundtruth item. MRR cares about the position of the groundtruth item and calculates the reciprocal of the rank at which the groundtruth item was put.

\begin{table*}
\caption{Performance comparison in terms of hit ratio and MRR on all datasets. Best performance is in boldface.  }
\label{result}
\begin{tabular}{c|c|ccccccccc|r}
\toprule
Dataset  & Metric & POP & BPRMF & MC  & FPMC& HRM & PRME & TransRec & Caser  & AttRec & Improv.\\
\toprule

\multirow{2}{*}{ML-100K}   & HR@50     & 0.2142 &0.3754 &   0.4115 & 0.4783 & 0.4821 &    0.4411 &  0.4634 & 0.4667 &  \textbf{0.5273} & 9.38\%  \\
& MRR   & 0.0388  &   0.0616 &  0.0662  & 0.0925  & 0.0889 & 0.0837 &   0.0827 &0.0799  &  \textbf{0.0981}  &10.35\%    \\
\midrule
\multirow{2}{*}{ML-HetRec} &   HR@50                      &  0.1065   &  0.1462   &   0.1903  &    0.2321 &       0.2380  &    0.2357   & 0.1912 &  0.2144    &  \textbf{0.2964}  &  24.54\%     \\
 &  MRR  &  0.0177   &  0.0215 &  0.0359  &  0.0489& 0.0486  &   0.0500   &   0.0337  &  0.0387 &  \textbf{ 0.0592}  & 18.40\%   \\
\midrule
\multirow{2}{*}{ML-1M}  &    HR@50    & 0.1440&  0.2378 &  0.3419&    0.4209  &  0.4311   & 0.4449  &  0.3358 &   0.4811    &    \textbf{0.5223}    & 8.56\%     \\
&   MRR   & 0.0231     & 0.0368    & 0.0654& 0.1022 &  0.0873  &  0.1044    &   0.0561 &  0.0925   &  \textbf{0.1172}   & 12.26\%   \\
\midrule

\multirow{2}{*}{Android App} &   HR@50 & 0.1194  &   0.1738  & 0.1802  & 0.1990 &  0.2001   &  0.1686& 0.2016  &  0.1426  &\textbf{ 0.2187} & 8.48\%\\
 & MRR  & 0.0228 &  0.0287  &   0.0355  & 0.0355  &  0.0295 &  0.0237  &0.0306  &  0.0231 &  \textbf{0.0415}   & 16.90\%  \\
\midrule

\multirow{2}{*}{Health / Care} &   HR@50 & 0.0337  &    0.0900 &0.0786   &  0.1128&     0.0965 & 0.0843 &0.0962   & 0.0768 &\textbf{ 0.1272} & 12.77\% \\
 & MRR  &0.0171  &0.0188   & 0.0245   &0.0258  & 0.0183 &0.0119    & 0.0232 &  0.0146 &  \textbf{0.0277}  & 7.36\%  \\
\midrule

\multirow{2}{*}{Video Game} &   HR@50 & 0.0609  &  0.1630   & 0.1708  & 0.2226 &  0.2150   & 0.1855 & 0.2035  &  0.1438  & \textbf{0.2414} & 8.45\% \\
 & MRR  & 0.0126&  0.0277  &   0.0381  &  0.0451 &  0.0337 &   0.0263 & 0.0349 &  0.0248 &   \textbf{0.0496}  & 9.98\%   \\
\midrule
\multirow{2}{*}{Tools / Home} &   HR@50 &  0.0319 &   0.0559 & 0.0384 & 0.0535& 0.0488 &  0.0465 & 0.0658& 0.0424   & \textbf{0.0775} & 17.78\% \\
 & MRR  & 0.0061 & 0.0099  &   0.0093  & 0.0129 & 0.0086& 0.0076&  0.0112  &  0.0071 &   \textbf{0.0148}  & 14.73\%  \\
\midrule
\multirow{2}{*}{Digital Music} &   HR@50 & 0.0436  &  0.1621   & 0.1307  & 0.1580 &  0.1998  &   0.1559 &  0.1894& 0.1327 &\textbf{0.2205} & 10.36\% \\
 & MRR  & 0.0073 &0.0277 & 0.0320   &  0.0322 & 0.0310  & 0.0243  &  0.0300 & 0.0228  &   \textbf{0.0467}  & 45.03\%   \\
\midrule

\multirow{2}{*}{Garden} &   HR@50 & 0.0319  &  0.0965   & 0.0946  & 0.1525 &  0.1593   & 0.1573 & 0.1486  & 0.1632   &\textbf{0.2177} &33.39\% \\
 & MRR  & 0.0049 &  0.0105&  0.0333  & 0.0408  &  0.0255 &  0.0266  & 0.0257 &0.0277  &   \textbf{0.0459}  & 12.50\%  \\
\midrule
\multirow{2}{*}{Instant Video} &   HR@50 & 0.1240 & 0.2350   & 0.1650 & 0.2120 &   0.2430  &  0.1910& 0.2570 &  0.1620  &\textbf{0.2790} &  8.56\% \\
 & MRR  & 0.0173 &  0.0376& 0.0426 &  0.0541 & 0.0414 &   0.0367 & 0.0441 &0.0275 &   \textbf{0.0634}   & 17.19\%  \\
\midrule
\multirow{2}{*}{LastFM}        & HR@50                        &     0.1314 & 0.3659 &   0.1682    &  0.2808 &     0.3733   & 0.2503    &     0.3785    & 0.1756 & \textbf{0.3901}    & 3.06\% \\
 & MRR      &0.0224  &0.1062 & 0.0645 &0.0869 &  0.1209 & 0.1276 &  0.1147 & 0.0343 & \textbf{0.1312}  & 2.82\%  \\
 \midrule
\multirow{2}{*}{MovieTweetings}        & HR@50                        &    0.1687   & 0.1749 &   0.3314    & 0.3417  &     0.3105  &   0.3286  &     0.2755  &  0.3332& \textbf{0.3602}   &   5.41\% \\
 & MRR      &  0.0204 & 0.0231& 0.0700 & 0.0674&   0.0534 &  0.0476 &  0.0421 &  0.0585& \textbf{0.0811} & 15.86\%  \\
\bottomrule
\end{tabular}
\end{table*}

%============================================
\subsection{Compared Models}
%============================================

Our model is dubbed as \textbf{AttRec} which can be considered as the abbreviation of ``attentive recommendation". We compare AttRec with classic methods as well as recent state-of-the-art models. Specifically, the following baseline models are evaluated.
\begin{enumerate}
    \item \textbf{POP}. This approach ranks the items based on their popularity in the system and the most popular items are recommended to users.
    \item \textbf{BPRMF}~\cite{rendle2009bpr}. It optimizes the matrix factorization in a pairwise manner with Bayesian Personalized Ranking loss, which aims to maximize the difference between positive and negative items. It does not model the sequential signals.
    \item \textbf{FMC}. This is a simplified version of factorized personalized Markov Chain (FPMC)~\cite{Rendle:2010:FPM:1772690.1772773} which does not include user personalized behaviours.

    \item \textbf{FPMC}~\cite{Rendle:2010:FPM:1772690.1772773}. This approach combines matrix factorization machine with Markov Chain for next item recommendations. The proposed approach  captures both user-item preferences and user sequential behaviours.
    \item \textbf{HRM}~\cite{Wang:2015:LHR:2766462.2767694}. It is a  Hierarchical Representation Model which captures both sequential and general user tastes by introducing both linear and nonlinear pooling operation for historical transaction aggregation. Here, the average aggregation is adopted.

    \item \textbf{PRME}~\cite{feng2015personalized}. This model was originally proposed for POI recommendation. It utilizes metric embedding to learn user and item embeddings as well as the user check-in sequences.

    \item \textbf{TransRec}~\cite{hetranslation,he2017translation}. This model applies the idea of translating embeddings~\cite{bordes2013translating} to sequential recommendation. It views users as relation vectors and assumes that the next item is determined by user's recent interacted item plus the user relation vectors.
    \item \textbf{Caser}~\cite{Tang:2018:PTS:3159652.3159656}. It models user past historical interactions with both hierarchical and vertical convolutional neural networks. It also considers the skip behaviors and the whole network is optimized by minimizing the cross entropy.
\end{enumerate}
Among all of these baselines, Caser and HRM are neural network based approach.  PRME and TransRec are metric embedding based algorithms. We omit comparisons with models such as Fossil~\cite{he2016fusing} and \textbf{GRU}~\cite{DBLP:journals/corr/HidasiKBT15} (\textit{RNN based sequential recommender}) since they have been outperformed by recently proposed Caser~\cite{Tang:2018:PTS:3159652.3159656} or TransRec~\cite{Tang:2018:PTS:3159652.3159656} model. Note that, in our experiments, we do not use pre-train for all models.

%============================================
\subsection{Implementation Details}
%============================================

The former seven baselines were implemented in C++ based on~\cite{he2017translation}. We implemented Caser and our model with Tensorflow\footnote{https://www.tensorflow.org/}. All experiments were conducted on a NVIDIA TITAN X Pascal GPU. For all baselines, Hyper-parameters are tuned with grid search with validation set.

Since we adopt adaptive gradient optimizer for AttRec, the learning rate of AttRec for all datasets is set to $0.05$ without further tuning. The number of latent dimensions $d$ of all latent vectors ($U, V, X$) of AttRec and all other baselines (if exists.) is set to $100$ for fair comparison. Note that the impact of $d$ is also discussed in the following section. Due to the high sparsity of Amazon, LastFm and Movietweetings datasets, we use norm clipping to replace the $\ell_2$ regularization for $X, V, U$. Weight matrices of nonlinear layer in self-attention module are regularized with $\ell_2$ norm. Regularization rate $\lambda$ is tuned amongst $\{0.1, 0.01, 0.001, 0.0001\}$. Dropout rate is tuned amongst $\{0, 0.3, 0.5, 0.7\}$. The weight factor $\omega$ is tuned amongst $\{0, 0.2, 0.4, 0.6, 0.8, 1.0\}$. The length of sequence $L$ is set to $5$ for Movielens, $3$ for MovieTweetings, and $2$ for all other datasets. The target length $T$ is set to $3$ for Movielens and $1$ for all other datasets.  The margin $\gamma$ of hinge loss is fixed to $0.5$ for all datasets.

%============================================
\subsection{Performance Comparison}
%============================================

Table~\ref{result} reports the experimental results of the 8 baselines and our model on 12 benchmark datasets. Observe that \textbf{AttRec} always achieve the best performance on all datasets. This ascertains the effectiveness of the proposed approach. Notably, the performance gains over the strongest baselines is reasonably large in terms of both prediction accuracy and ranking quality. Our model performs well not only on dense datasets like Movielens but also on sparse datasets such as Amazon or MovieTweetings. The sequential intensity of sparse datasets is usually much lower than that of dense datasets.

%The Caser model even removed comparison on sparse datasets due to its weaker sequential signals~\cite{Tang:2018:PTS:3159652.3159656}. However, AttRec is capable of dealing with both cases.
Additionally, we make several observations on the comparison baselines. Markov Chain based models (FPMC and MC) achieve consistent performance on both dense and sparse data. TransRec and PRME, on the other hand, seems to be underperforming on some datasets. One important assumption of PRME and TransRec is that user's next item is only influenced by her latest action. This assumption might hold on sparse data as the interactions are extremely discrete along time but may not hold when user interacts with the system frequently. TransRec overcomes this shortcoming to some extent by introducing user specific relation vectors as intermediary. This claim can be demonstrated by HRM as it usually outperforms PRME, and their is no clear winner between HRM and TransRec.  Caser achieves satisfactory performance on Movielens and MovieTweetings, but  performs poorly on sparse datasets. As a final recapitulation, AttRec consistently outperforms all baselines by a wide margin, which clearly answers \textbf{RQ1}.

%============================================
\section{Model Analysis and Discussion}
%============================================

In this section, we dive into an in-depth model analysis, aiming to further understand behaviour of our model to answer the \textbf{RQ2}.

\begin{table}
\caption{HR@50 of AttRec with and without Self-Attention.}
\label{wandwoatt}
\begin{tabular}{c|cccc}
\toprule
Dataset            & ML-100K & ML-1M & Garden & Digit Music \\ \midrule
w/ Self-Att  & 0.5273  & 0.5223    & 0.2177 & 0.2205        \\
w/o Self-Att &  0.5015      &   0.5045      &    0.2026    &         0.2022     \\
\bottomrule
\end{tabular}
\end{table}

%============================================
\paratitle{Impact of Self-Attention.} Although we can infer the efficacy of self-attention implicitly from Table~\ref{result}, we would like to verify the effectiveness of the self-attention mechanism explicitly. We remove the self-attention module from AttRec and replace $m^u_t$ with the mean of $X^u_{t}$, that is: $m^u_t = \frac{1}{L} \sum^{L}_{l=1} X^u_{tl}$.

\begin{figure}
\subfloat[User ID: 538 (Male)]{\includegraphics[width=.49\columnwidth]{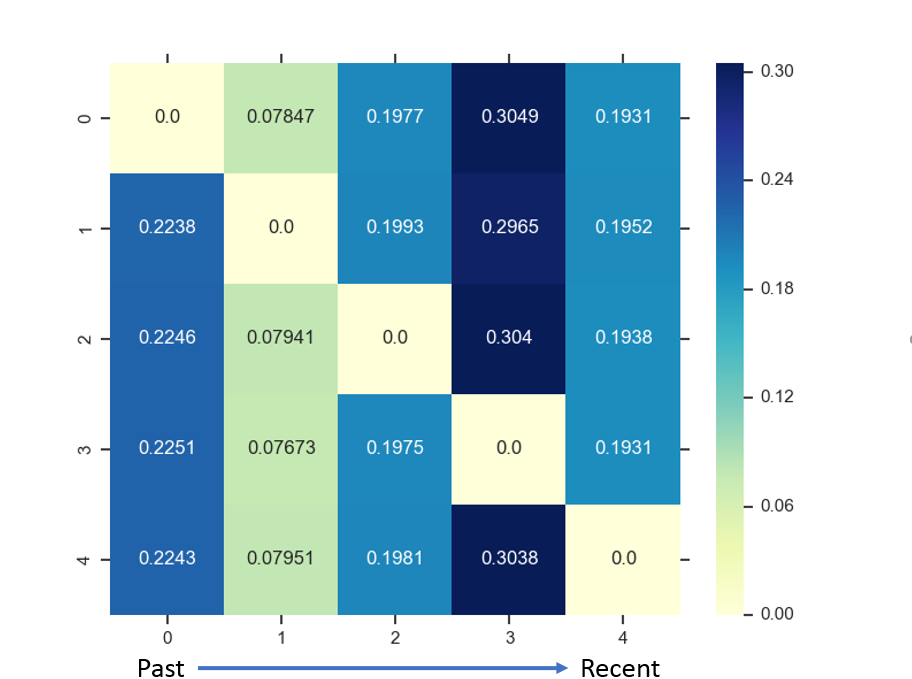}}
\subfloat[User ID: 888 (Male)]{\includegraphics[width=.49\columnwidth]{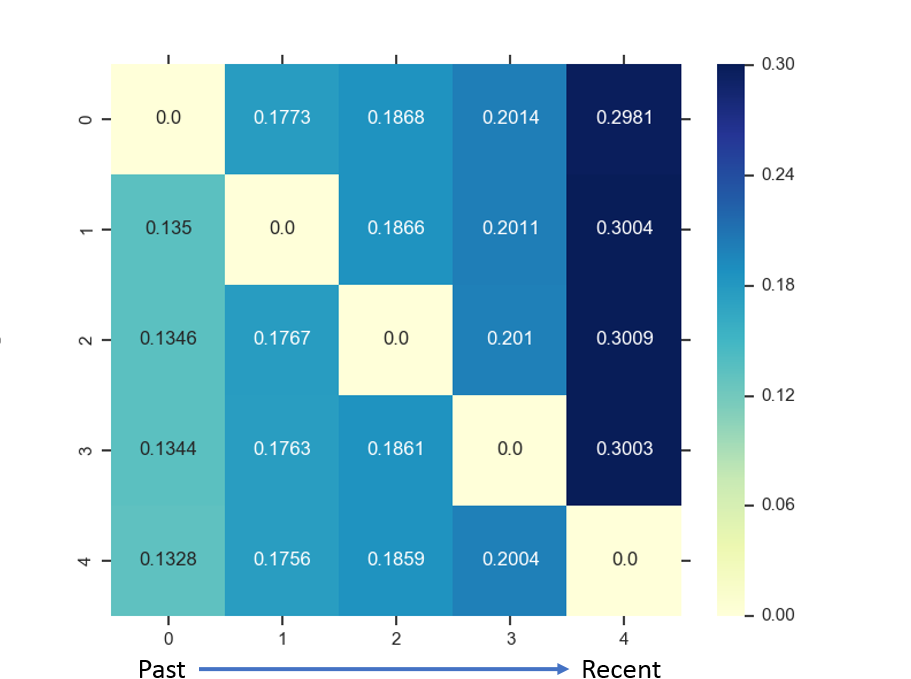}}
\caption{Attention weights of two randomly selected users for  prediction. The color scale represents the strength of the attention weights. Item information is not listed due to limited space. (\textit{Best viewed in color.})}
\label{fig:attentionmap}
\end{figure}

Table~\ref{wandwoatt} shows the comparison between with and without self-attention. We observe that with self-attention indeed improves the performance. From both Tables~\ref{result} and Table~\ref{wandwoatt}, we find that even without self-attention, our model can still beat all baselines on these four datasets. This also justifies the method we use for preference modelling. Furthermore, in order to study the effect of self-attention, we visualize the self-attention matrix on Movielens 100K in Figure~\ref{fig:attentionmap}.

We make two observations from the results. First, the attention matrix is column distinguishable even they are unintentionally trained to achieve this.  Each column represents the importance and weight for the corresponding action. Intuitively, the self-attention matrix allows us to inspect how much an action contributes to the overall short-term intent representation. Second, AttRec will not simply put more emphasis on recent actions, but learns to attend over previous actions with different weights automatically. For example, the most recent items are given higher weights for user ``888", while higher weights are assigned to the first and fourth items for user ``538". Evidently, analyzing the attention weights could lead to benefits in interpretability.

\begin{table}
\centering
\caption{HR@50 of AttRec with different aggregation methods.}
\label{aggre}
\begin{tabular}{c|cccc}
\toprule
Dataset & ML-100K & ML-1M & Garden & Digit Music  \\ \midrule
Mean    & 0.5273  & 0.5223    & 0.2177 & 0.2205        \\
Sum     &   0.4883      &   0.5201        &   0.2046     &        0.1908       \\
Max     &   0.5254      &   0.5229        &    0.1892    &    0.1925           \\
Min     &   0.5244      &     0.5267      &     0.1525   &       0.1548        \\
\bottomrule
\end{tabular}
\end{table}

%============================================
\paratitle{Impact of Aggregation Method.} As aforementioned,  we can use different aggregation strategies to get the representation of user short-term intents. Here, we explored four types of strategies to check their suitability. Table~\ref{aggre} shows the results of using different aggregation methods. We observe that ``mean" achieves desirable performance on both sparse and dense datasets. The other three aggregation methods seem to be underperforming especially on sparse datasets. This is reasonable as $m^u_t$ shall influence the embedding of the next item ($X^u_{t+1}$). Using mean aggregation could retain more information.

\begin{figure}
\centering
\subfloat[Effects of the weight $\omega$\label{sfig:paraw}]{\includegraphics[width=.48\columnwidth]{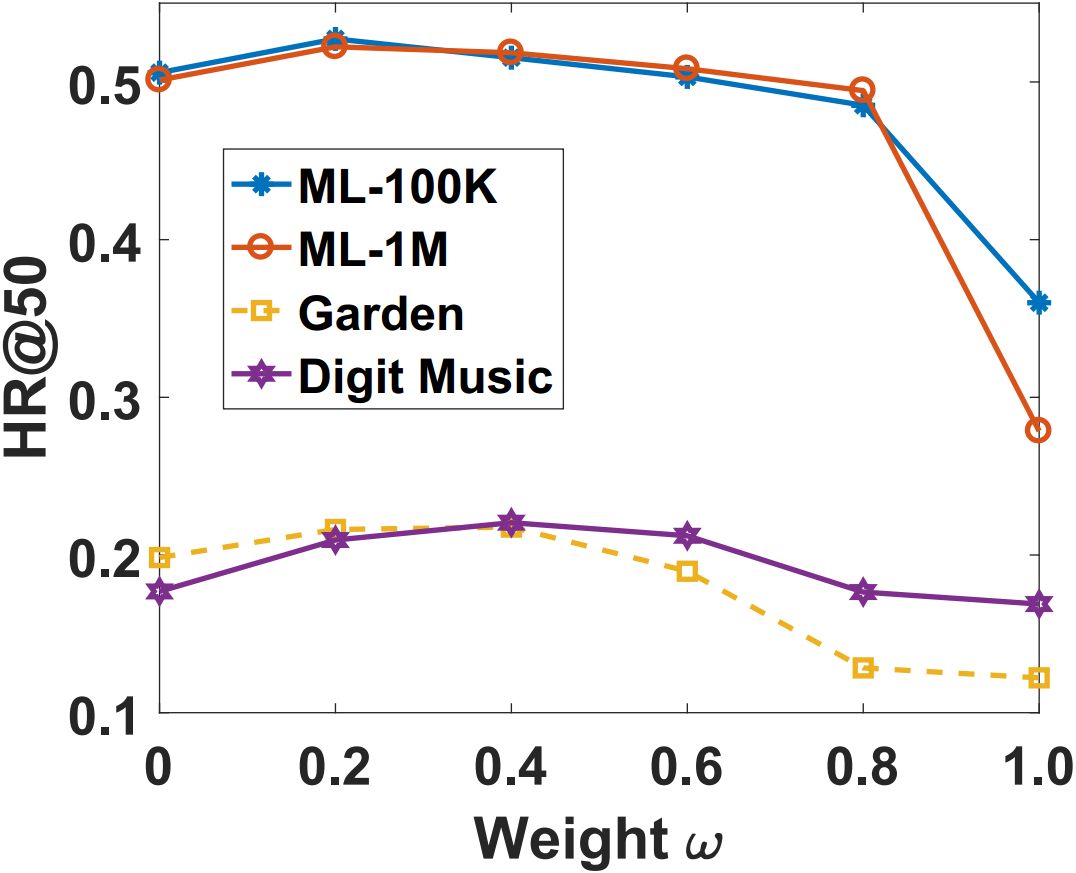}}
\quad
\subfloat[Effects of the sequence length $L$\label{sfig:paral}]{\includegraphics[width=.48\columnwidth]{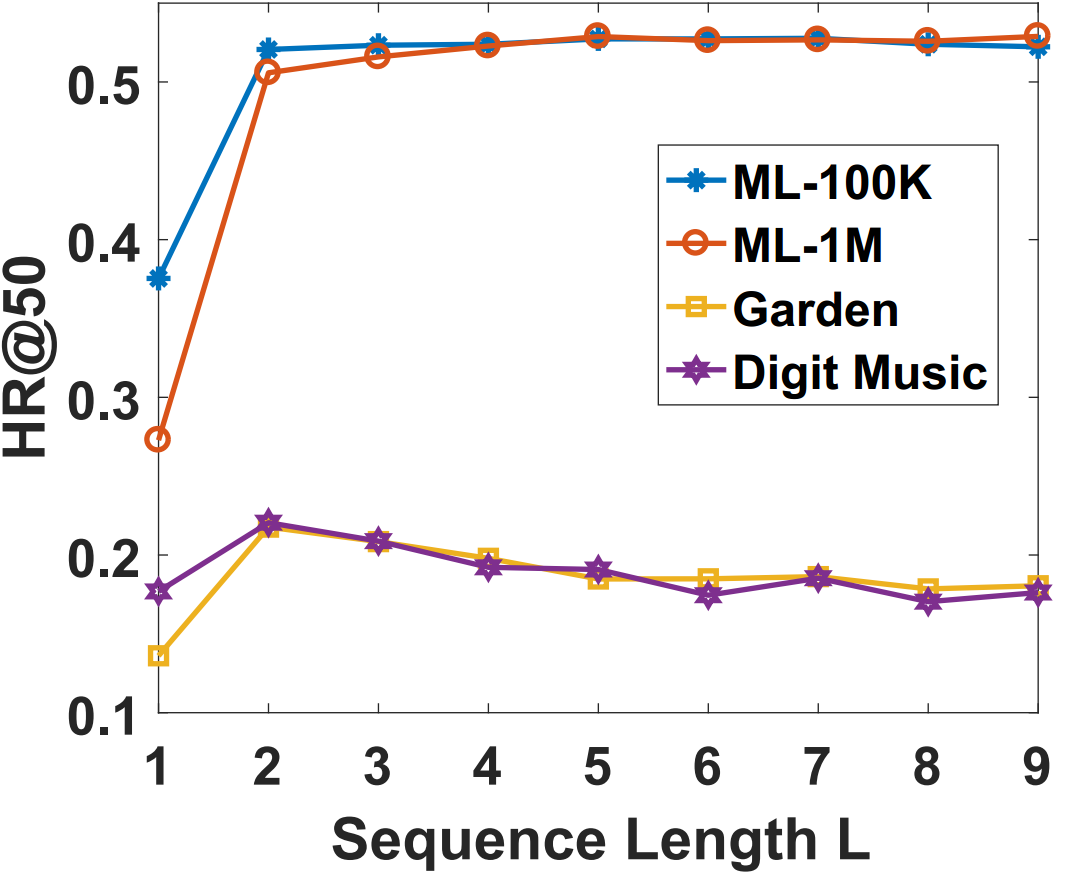}}
\caption{ Effects of the weight $\omega$ and effects of the sequence length $L$ on four datasets. }
\label{fig:parame}
\end{figure}

%============================================
\paratitle{Impact of weight $\omega$.} The parameter $\omega$ controls the contribution of short-term and long-term effects. Observe from Figure~\ref{sfig:paraw},  considering only short term intents ($\omega=0$) usually get better performance than considering only long-term preference ($\omega=1.0$). Setting $\omega$ to a value between $0.2$ and $0.4$ is more preferable, which also indicates that short-term intent play a more important role in sequential recommendation. Additionally, the impact of $\omega$ also reflects the strength of sequential signal in the datasets. Datasets with higher sequential signal (dense datasets such as Movielens, detail sequential intensity evaluation can be found in~\cite{Tang:2018:PTS:3159652.3159656}) hit their best performance with a lower $\omega$ value.

%============================================
\paratitle{Impact of sequence length $L$.} Figure~\ref{sfig:paral} shows the impact of the sequence length $L$. We observe that the proper $L$ is highly dependent on the density of datasets. On MovieLens datasets where average number of actions per user is greater than a hundred, setting $L$ to a larger value is beneficial to the performance. However, $L$ should be set to a small value on sparse datasets, which is reasonable as increasing $L$ will results in training sample decrease. Note that self-attention is capable of drawing dependencies between distant positions~\cite{vaswani2017attention}, which theoretically allows learning on very lengthy sequence.

\begin{figure}
\centering
\includegraphics[width=7.0cm]{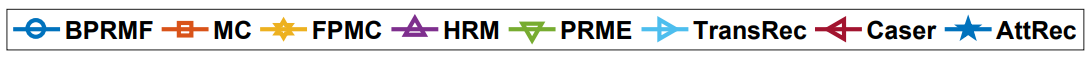}\\
\subfloat[Movielens 100K]{\includegraphics[width=.48\columnwidth]{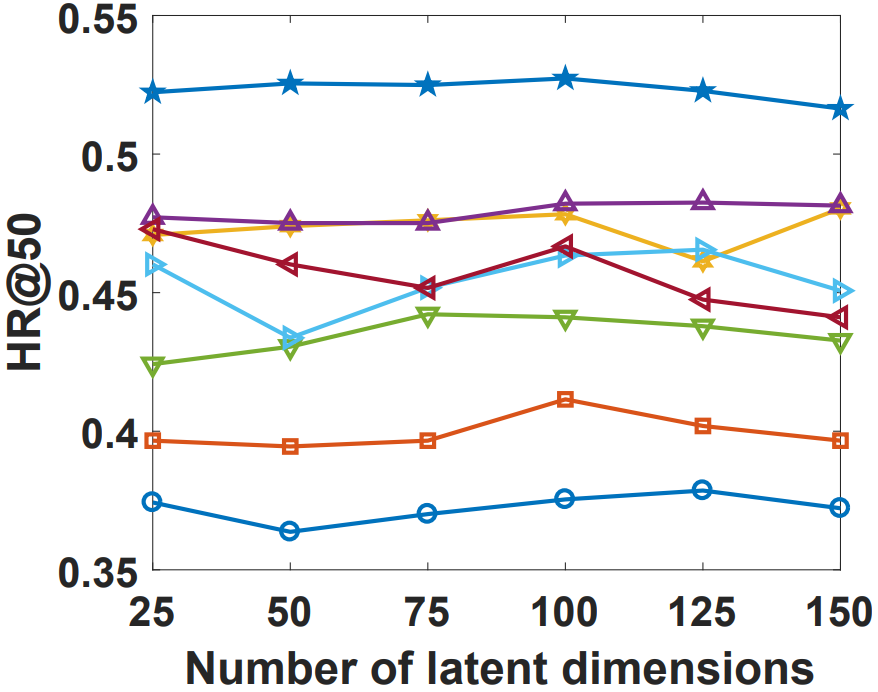}}
\quad
\subfloat[Amazon Garden]{\includegraphics[width=.48\columnwidth]{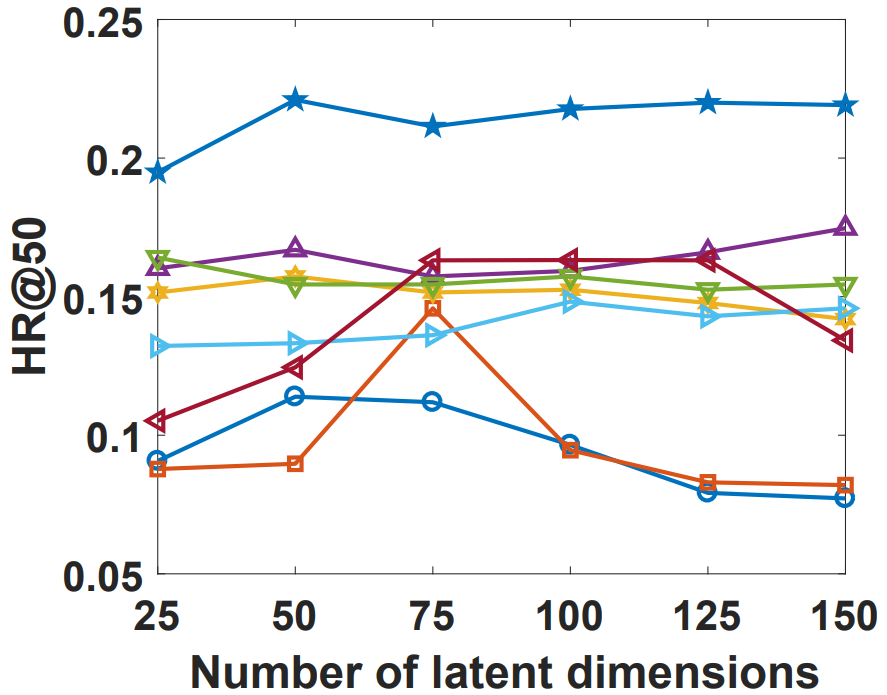}}
\caption{HR@50 (y-axis) vs. the number of latent dimensions (x-axis) on Movielens 100K and Amazon Garden.}
\label{fig:latent}
\end{figure}

%============================================
\paratitle{Impact of Number of Latent Dimensions.} Figure~\ref{fig:latent} shows the HR@50 for various $d$ while keeping other optimal hyper-parameters unchanged. We make three observations from this figure. First, our model consistently outperforms all other baselines on all latent dimensions. Secondly, a larger latent dimension does not necessarily leads to better model performance. Overfitting could be a possible reason. Third, some models such as MC and Caser perform unstably, which might limit their usefulness.

%============================================
\paratitle{Model Efficiency.} Table~\ref{runtime} shows the runtime comparison with Caser. Other baselines are not listed here as the implementation cannot leverage the computation power of GPU. Experiments were run with batch size of $1000$ on a NVIDIA TITAN X Pascal GPU.  We observe that AttRec only incurs a small computational cost over Caser. This cost might be caused by the use of dual embedding and Euclidean distance calculation. Since both Caser and AttRec are trained in a pairwise manner, the difference in convergence speed is subtle. For example, it takes fewer than $30$ epochs for AttRec to achieve its best performance on Movielens 100K.

\begin{table}
\centering
\caption{Runtime comparison (second/epoch) between AttRec and Caser on four datasets.}
\label{runtime}
\begin{tabular}{c|cccc}
\toprule
Dataset            & ML-100K & ML-1M & Garden & Digit Music \\ \midrule
AttRec  &  1.315&  15.429   &  0.169 & 1.525        \\
Caser &   1.309    &   15.328    &    0.120    &         0.956      \\
\bottomrule
\end{tabular}
\end{table}

%============================================
\section{Conclusion}
%============================================

In this paper, we proposed AttRec, a novel self-attention based metric learning approach for sequential recommendation. Our model incorporates both user short-term intent and long-term preference to predict her next actions. It utilizes the self-attention to learn user's short-term intents from her recent actions. Analysis shows that AttRec could accurately capture the importance of user recent actions. In addition, we generalize self attention mechanism into metric learning methodology for sequence prediction task, which possess good effectiveness on sequential recommendation.

In the future, we will investigate incorporating user and item side information to overcome the sparsity problem. Additionally, we believe that our model is adaptable to other related sequence prediction tasks, such as session-based recommendation or click-through prediction.

\paratitle{Acknowledgment}. To be completed in final version.

\bibliographystyle{ACM-Reference-Format}
\bibliography{SeqRecBib}

\end{document}